\documentclass[prd,twocolumn,nofootinbib,preprintnumbers,superscriptaddress]{revtex4-1}

\usepackage{times}
\usepackage{natbib}
\usepackage{amssymb,amsbsy,amsmath,amsfonts}
\usepackage{mathrsfs}
\usepackage{graphicx}% Include figure files
\usepackage{epsf,epsfig,float,latexsym,amsthm,fancyhdr,rotating}
\usepackage{graphics,psfrag,longtable}
\usepackage{slashed}
\usepackage{esint}
\usepackage{nicefrac}
\usepackage{braket}
\usepackage{mathtools}

\newcommand{\be}{\begin{equation}}
\newcommand{\ee}{\end{equation}}

\usepackage[colorlinks,citecolor=blue,linktoc=all,linkcolor=cyan]{hyperref}

\begin{document}
\title{Soft-Photon Contribution into Two-Photon Exchange Corrections for Azimuthal Asymmetries of SIDIS}

\author{Stinson Lee}

\author{Andrei Afanasev}

\affiliation{Department of Physics,
The George Washington University, Washington, DC 20052, USA}

\begin{abstract}
		It is demonstrated that two-photon exchange (TPE) corrections to the cross-section of unpolarized semi-inclusive deep-inelastic scattering (SIDIS) generate azimuthal-dependent terms and the corresponding $\braket{{\cos{(n\phi)}}}$ moments. A quark-diquark model of a nucleon was used in the calculations along with a soft-photon approximation. The infrared divergences in the intermediate steps of calculation are regularized with the fictitious photon mass that cancels out in the final result when the soft-photon bremsstrahlung process and interference terms are added.  The calculations employ Mathematica ``LoopTools" package to evaluate the loop integrals. TPE corrections are analyzed in the kinematics of planned experiments at Jefferson Lab.
		\end{abstract} 
\maketitle

\section{Introduction}

Semi-Inclusive Deep Inelastic Scattering (SIDIS) is a powerful tool for probing the internal structure of nucleons at the partonic level, allowing for 3D momentum tomography of hadrons \cite{achenbach2024present, achenbach2024present, aschenauer2019semi, accardi2016electron, accardi2024strong}. In SIDIS, a high-energy lepton scatters off a nucleon, exchanging a virtual photon, which subsequently interacts with the partonic constituents - the charged quarks - within the nucleon. The scattered lepton and at least one of the produced hadrons are detected in the final state, providing access to multidimensional distributions of quarks inside the nucleon. Using a one-photon exchange approximation in QED, the differential cross-section of SIDIS is expressed in terms of well-defined kinematic variables. Studies of kinematic dependence enable the extraction of transverse momentum-dependent distributions (TMDs), which encode vital information on the motion of quarks inside hadrons, that is a focus of current and future experimental programs at particle accelerators \cite{achenbach2024present, aschenauer2019semi, accardi2016electron, accardi2024strong}.
%\cite{boer2001sudakov}.

An important aspect of electron-hadron scattering that gained attention in recent years is the role of two-photon exchange (TPE) corrections; see Refs.\cite{afanasev2017two, afanasev2024radiative} for review. The TPE mechanism involves the exchange of two virtual photons between the charged lepton and the nucleon, introducing higher-order QED corrections that can alter kinematic dependence of the measured cross sections due to additions that cannot be expressed in terms of form factors or structure functions defined within a one-photon exchange approximation. Although TPE contributions are generally suppressed in comparison to single-photon exchange, they can become significant in certain kinematic regions, particularly where the physical effects studied require relative accuracy of a few per cent or better. Theoretical studies for exclusive electroproduction of pions on nucleons \cite{afanasev&aleksejevs&barkanova,cao2020two} suggest that TPE effects introduce modifications to the cross-section that go beyond conventional radiative corrections \cite{akushevich1999radiative,akushevich2009lowest}, impacting the extraction of generalized parton distributions in deep-exclusive meson production experiments.

In this article, we extend the assessment of the role of TPE to SIDIS processes using the framework of a quark-diquark model and a soft-photon exchange approximation in which one of the exchanged photons is treated in the zero-momentum limit. This effective model, which simplifies the nucleon as a system consisting of an active quark and a spectator diquark, provides a tractable approach for studying partonic interactions \cite{brodsky&hwang&schmidt,afanasev&carlson}. The soft-photon-exchange approach allows us to identify a model-independent part of TPE corrections in SIDIS that is defined only by the general parameters of the reaction such as  charges, masses and kinematic variables of the particles involved, but is independent of underlying strong dynamics of the hadrons.  The results presented here offer insights into the phenomenological consequences of TPE in SIDIS and contribute to a more comprehensive understanding of nucleon structure.

\section{Two-Photon Exchange for SIDIS}

\begin{figure}[hbt!]
    \centering
	\includegraphics[scale=0.08]{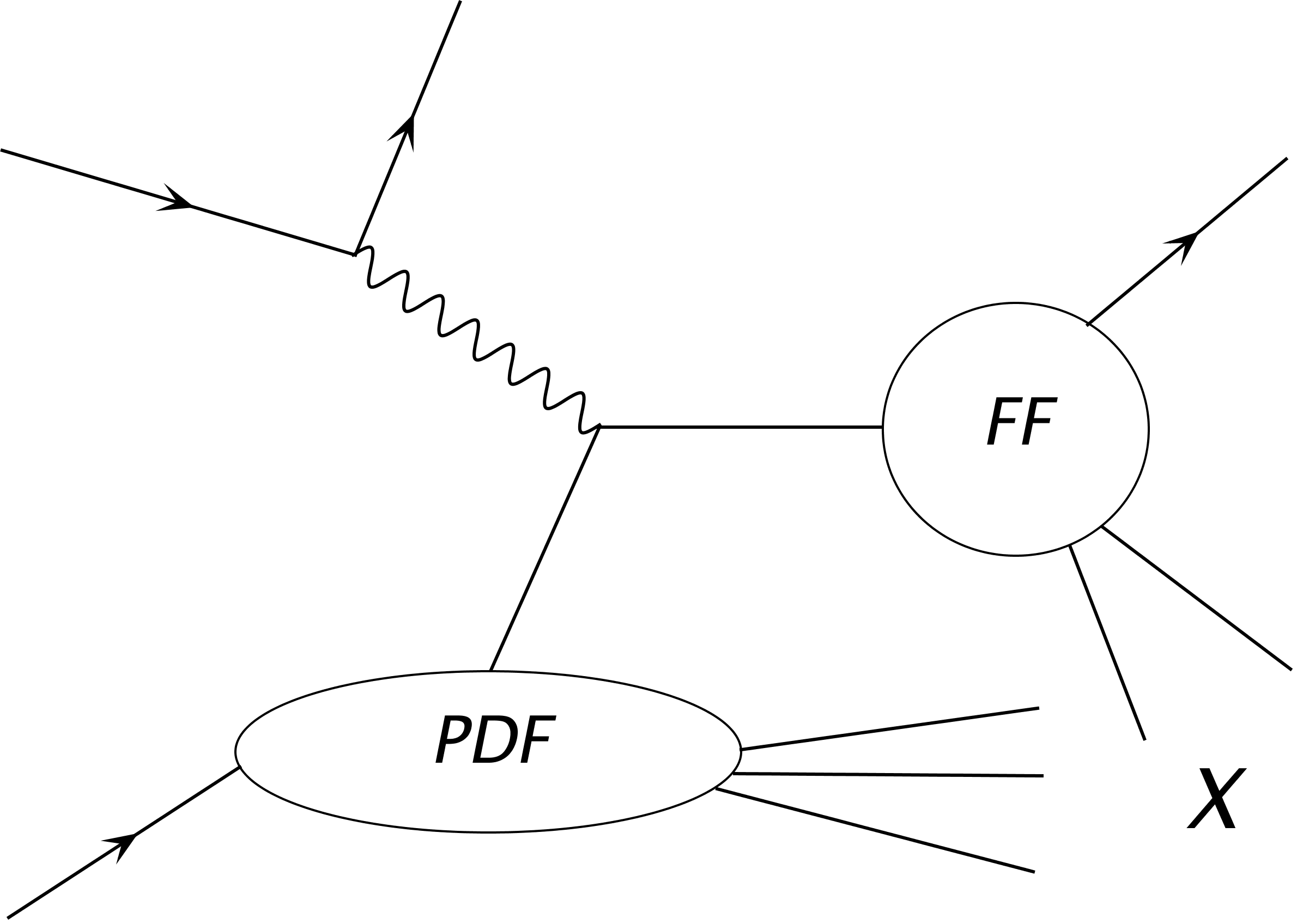}
	\caption{\small{Single-photon exchange in the SIDIS process, illustrating the involvement of PDFs and FFs. X represents the spectator system.}}
	\label{fig:parton}
\end{figure}

The parton model for SIDIS (Fig.\;\ref{fig:parton}) provides a schematic representation of the scattering process in a one-photon exchange approximation, with fragmentation functions (FF) describing formation of a hadron from a struck parton, while parton distribution functions (PDF) are defined by intrinsic motion of partons within the nucleon. 
This internal motion significantly influences the kinematic distributions of the final-state hadrons observed in the SIDIS experiments. Unlike collinear factorization, which assumes that partons carry the entire nucleon momentum along its longitudinal direction, TMD factorization explicitly incorporates the transverse momentum of partons \cite{bacchetta&boer&diehl&mulders, collins2011foundations, collins2017connecting}. 

TPE introduces systematic corrections to the scattering cross sections, evaluation of which requires a well-defined model of nucleon structure. Fortunately, a significant part of TPE amplitude can be treated within the framework of Low-Energy Theorems (LETs) \cite{low1954scattering, gell1954scattering,low1958bremsstrahlung, saito1969low}, which emphasize the collective properties of the nucleon rather than the resolution of individual quark constituents. In the context of TPE calculations, a soft-photon exchange approach based on LET was implemented in Ref.\cite{tsai1961radiative, maximon2000radiative} for elastic electron-proton scattering and in Ref.\cite{afanasev&aleksejevs&barkanova} for the exclusive electroproduction of pions.
 %These theorems provide valuable insights into scattering phenomena in regimes where standard perturbative methods become unreliable, highlighting the necessity of incorporating non-perturbative effects. 
In the following sections, we focus on the soft-photon exchange approach, which is model-independent and adheres to LETs, highlighting the significance of TPE effects. 

For a more complete theoretical description, it is necessary to also consider the exchange of two hard photons, which involves calculations at the quark level. Hard-photon exchange may be added independently to the soft-photon contribution considered here, and this task is deferred to a future publication. 
%This approach offers a more refined perspective on the scattering dynamics. Within the TMD framework, parton distribution and fragmentation functions play a central role in describing both the momentum and spatial distributions of partons inside the nucleon. By carefully analyzing the kinematics of scattered leptons and produced hadrons in the SIDIS experiments, researchers can extract detailed information on the parton dynamics.

%Beyond its relevance to TMD studies, TPE also provides a unique means of probing the electromagnetic structure of the nucleon. The exchange of two virtual photons between the incoming lepton and the nucleon target introduces subtle modifications to the scattering cross-section, which can lead to measurable corrections that impact the interpretation of experimental results. A thorough theoretical investigation of TPE effects in the TMD framework enhances our understanding of hadronic structure and the fundamental mechanisms governing SIDIS. Moreover, recognizing the impact of TPE in SIDIS is essential to refining our knowledge of the nucleon electromagnetic properties and improving the precision of phenomenological models.

\subsection{Amplitudes and TPE Correction}
In order to evaluate the effect of TPE on SIDIS observables, we used a quark-diquark model \cite{brodsky&hwang&schmidt, afanasev&carlson} that proved efficient in demonstrating novel effects in SIDIS. This model assumes a simplified representation of the proton as a bound state of a quark and a scalar diquark.
%During the TPE process, diquarks couple to external photons through the electric charges of their constituent quarks, influencing the polarization observables. 
In a quark-diquark model describing an electron scattering off a nucleon, where the final state includes a spin-1/2 quark ($q{\prime}$) and a scalar diquark ($S$), the interaction can be represented as
\begin{center}
    $e(k_1) + N(k_2)\rightarrow e(k_3)+ q'(k_4) + S(k_5)$,
\end{center}
%This interaction exhibits similarities with pion electroproduction, allowing for the experimental detection of the quark jet in the final state. 

\begin{figure*}[hbt!]
	\centering
	\includegraphics[scale=0.6]{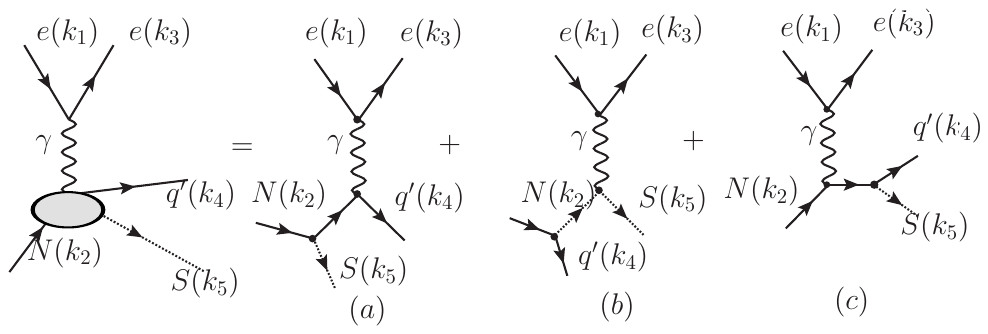}
	\vspace{-40em}
	\caption{\small{Reaction amplitude in one-photon approximation within a quark-diquark model. (a) is "quark graph", (b) is "diquark graph", and (c) is "proton pole graph"; $q'$ and $S$ stand for quark and diquark, c.f. Ref.~ \cite{afanasev&carlson}. }}
	\label{fig:1gamma}
\end{figure*}

 The Born level amplitudes, as shown in Fig.\;\ref{fig:1gamma}, $M^{1\gamma}=M_a^{1\gamma}+M_b^{1\gamma}+M_c^{1\gamma}$ with the subscripts a,b,c represent the ``quark" (Fig.\;\ref{fig:1gamma}a), the ``diquark" (Fig.\;\ref{fig:1gamma}b), and the ``proton pole" (Fig.\;\ref{fig:1gamma}c). The matrix elements for the nucleon-quark (diquark) part are $J^\nu\equiv J^{1\gamma}=J_a^{1\gamma}+J_b^{1\gamma}+J_c^{1\gamma}=\Bar{u}(k_{4(5)},m_{4(5)})\Gamma_\alpha^{N-\gamma-q'-S}u(k_2,m_2)$ with
\begin{align}
\label{eq:Born1}
\begin{split}
    J_a^{1\gamma}=&-\frac{e_{q'}g}{(k_4-q)^2-m_4^2}\Bar{u}(k_4,m_4)\gamma^\mu\\&\cdot(\slashed{k}_4-\slashed{q}+m_4)u(k_2,m_2)
\end{split}
\end{align}
\begin{align}
\label{eq:Born2}
\begin{split}
    J_b^{1\gamma}=&-\frac{e_rg}{(k_5-q)^2-m_5^2}\Bar{u}(k_5,m_5)\\&\cdot(2k_5^\mu-q^\mu)u(k_2,m_2)
\end{split}
\end{align}
\begin{align}
\label{eq:Born3}
\begin{split}
    J_c^{1\gamma}=&-\frac{e_Ng}{(k_2+q)^2-m_2^2}\Bar{u}(k_4,m_4)\gamma^\mu\\&\cdot(\slashed{k}_2+\slashed{q}+m_2)u(k_2,m_2),
\end{split}
\end{align}
where notations for particles' 4-momenta are shown in in Fig.\ref{fig:1gamma},  $u(k_2,m_2)$ and $u(k_4,m_4)$ denote bi-spinors of the target proton and a final quark, respectively. 
%Further, photon's coupling to the diquark can be ignored for production of a charged hadron.
The matrix elements can be written with the choices for the charges $e_l=e_N,e_s,$ or $e_{q'}$ for different cases (quark graph (a), diquark graph (b), proton pole graph (c)) with different charges: $e$ is the electron charge, $e_{q'}$ is the quark charge, the charge of the proton is $e_N=e_{q'}+e_s=1$, and $e_s$ is the charge of the diquark \cite{afanasev&carlson}.

Next we consider TPE contributions within the same model. The associated Feynman diagrams, illustrated in Figs.\;\ref{fig:two-photon},\ref{fig:bremsstrahlung}, decompose into two principal contributions: (i) TPE (Fig.\;\ref{fig:two-photon}) and (ii) (soft-photon) bremsstrahlung (Fig.\;\ref{fig:bremsstrahlung}). Inclusion of bremsstrahlung is important because interference between bremsstrahlung from electron and hadron lines of (Fig.\;\ref{fig:bremsstrahlung}) provides for cancellation of infrared (IR) divergence of TPE contributions of Fig.\;\ref{fig:two-photon}; only their combined effect is IR-finite. %Maintaining gauge invariance is essential to ensure the physical consistency of these calculations, as has previously been discussed in theoretical studies \cite{afanasev&carlson}. At the lowest order, electron scattering within the quark-diquark framework can be described by three fundamental diagrams as shown in Fig.\;\ref{fig:1gamma}: the quark graph, the diquark graph, and the proton pole graph. 

%Overall, the corresponding matrix element for the electroproduction process is formulated as:
%\begin{equation}
%    M\sim \braket{e(k_3)|j_\nu|e(k_1)}\times\braket{q'(k_4)S(k_5)|J^\nu|N(k_2)}
%\end{equation}
%where $j_\nu$ and $J^\nu$ represent the respective electron and hadron electromagnetic current operators mediating the interaction.

\subsection{TPE in Soft-Photon Approximation}

\begin{figure}[h]
\begin{minipage}[]{0.9\linewidth}
    \centering
    \includegraphics[width=0.8\linewidth]{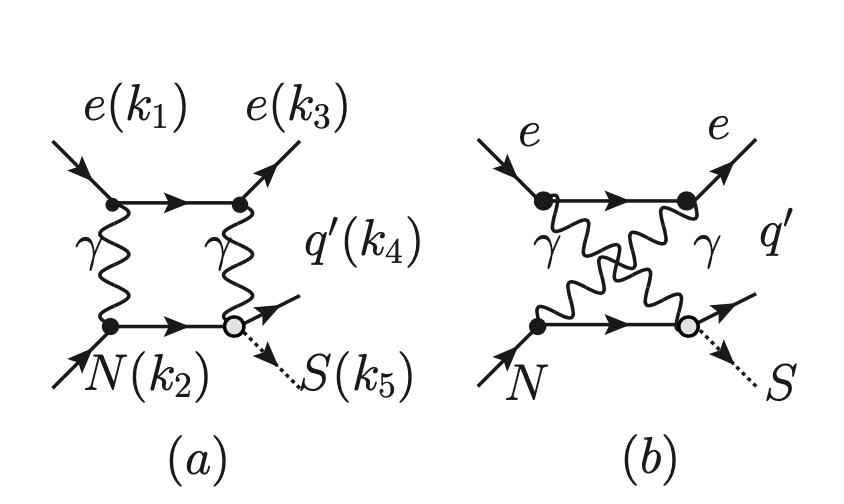}\\
    \label{}
\end{minipage}
\\
\vspace{0.2 cm}
\begin{minipage}[]{0.9\linewidth}
    \centering
    \includegraphics[width=0.8\linewidth]{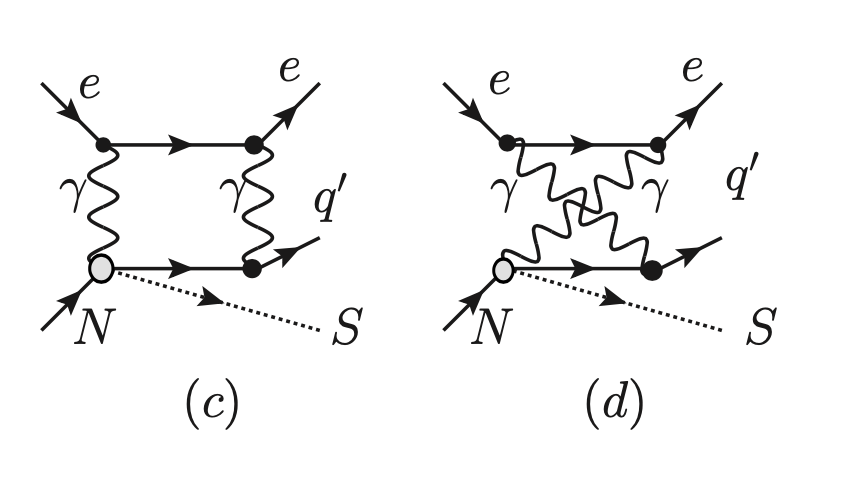}\\
    \label{}
\end{minipage}
\caption{\small{TPE diagrams within a quark-diquark model, where $q'$ represents a struck quark, or any other detected hadron in the respective process, and $S$ represents a diquark, or the spectator of the process.}}
\label{fig:two-photon} 
\end{figure}

Using the soft-photon approximation, which is neglecting the momentum for one of the virtual photons during the calculations, the Born-level amplitude for box diagrams as Fig.\;\ref{fig:two-photon} shown can be expressed as the Passarino-Veltman three-point scalar integral. This is similar to a previous study \cite{afanasev&aleksejevs&barkanova} that presented the calculation of the TPE for exclusive pion electroproduction.

%The Born-level amplitude expressed as
%\begin{equation}
%    \begin{split}
%        M^{1\gamma}&=[\Bar{u}(k_3,m_e)i\gamma_\mu u(k_1,m_e)]\cdot[\Bar{v}(k_{4(5)},m_{q'(s)})\Gamma_\alpha v(k_2,m_p)]\\
%        &\cdot\frac{g^{\mu\alpha}}{(k_3-k_1)^2},
%    \end{split}\label{eqn:born-M0}
%\end{equation}
%and it can lead to the amplitude for TPE that is represented as Eqn.\;\ref{eqn:M-box-spt} and Fig.\;\ref{fig:two-photon}a. 

Using Tsai's soft-photon approximation  (SPT) \cite{tsai1961radiative}, which neglects one of the photons' momenta in TPE amplitude, the photon propagator is approximated as:
$\frac{g^{\mu\alpha}}{(q+k_3-k_1)^2}\rightarrow\frac{g^{\mu\alpha}}{(k_3-k_1)^2}$.
Using Dirac algebra together with SPT prescription, terms like $(\slashed{k}_i+m_i)\gamma_\nu u(k_i,m_i)$ can be simplifed into $2k_{i\nu} u(k_i, m_i)$.
%The photon-proton coupling $\Gamma_\beta^{p-\gamma-p}=-ie\gamma_\beta$ under the SPT prescription and the pion photoproduction coupling $\Gamma_\alpha$ do not contribute to the radiative corrections in the soft-photon approximation. 
Other cases can be expressed in a similar format with the corresponding momentum terms $k_j$.

The box diagram amplitude (Fig.\;\ref{fig:two-photon}) under the SPT prescription is given as
\begin{equation}\label{eqn:M-box-spt}
    \begin{aligned}
        M^{2\gamma}&=\frac{i}{16\pi^4}\int dq^4[\Bar{u}(k_3,m_e)ie\gamma_\mu\frac{\slashed{k}_i\mp\slashed{q}+m_i}{(k_i\mp q)^2-m_i^2}ie_l\cdot\\
        &\gamma_\nu u(k_1,m_e)]\cdot[\Bar{u}(k_4,m_{q'})(\Gamma_\alpha^{p-\gamma-\pi-n})\frac{\slashed{k}_j\pm\slashed{q}+m_j}{(k_j\pm q)^2-m_j^2}\\
        &\cdot\Gamma_\beta^{p-\gamma-p} u(k_2,m_p)]\cdot\frac{g^{\nu\beta}}{q^2}\cdot\frac{g^{\mu\alpha}}{(k_3-k_1+q)^2}.
     \end{aligned}
\end{equation}
The Born-level amplitude follows from Eqs.(\ref{eq:Born1}-\ref{eq:Born3}):
\begin{eqnarray}
        M^{1\gamma}=[\Bar{u}(k_3,m_e)i\gamma_\mu u(k_1,m_e)]\\ \nonumber \cdot[\Bar{u}(k_{4(5)},m_{q'(s)})\Gamma_\alpha u(k_2,m_p)]
        &\frac{g^{\mu\alpha}}{(k_3-k_1)^2}.
        \label{eqn:born-M0}
\end{eqnarray}
The main outcome of the soft-photon approximation is that the amplitude of the box diagram reduces to the amplitude in one-photon exchange approximation times the factors that only depend of kinematics of the external particles:
\begin{align}
\begin{split}
    M^{2\gamma}=&M^{1\gamma}\cdot\sum_l[\frac{-ee_l}{2\pi}\cdot \sum_{i,j}(2k_i\cdot k_j) \\&\cdot C_0(\{k_i,m_i\},\{\mp k_j,m_j\})
\end{split}\\
    \equiv&\sum_{l=N,q',s}\sum_{i=a,b,c}M^{1\gamma}M_{l,i,box},
\end{align}
where the Passarino-Veltman three-point scalar integral \cite{passarino1979one} is defined as
\begin{align}
\label{eqn:scalar-integral}
    \begin{split}
        &C_0(\{k_i,m_i\},\{k_j,m_j\})=\\&\frac{1}{i\pi^2}\int d^4q\frac{1}{q^4}\cdot\frac{1}{(k_i-q)^2-m_i^2}\cdot\frac{1}{(k_j-q)^2-m_j^2} .
\end{split}
\end{align}

The correction due to TPE becomes 
\begin{equation}
    \delta_{box}^{SPT}=\frac{2Re[M^{2\gamma}M^{1\gamma\dagger}]}{|M^{1\gamma}|^2}=2Re[\sum_{l;i} M_{l,i,box}],
\end{equation}
$i.e.$ it is independent of the model used to describe a one-photon-exchange approximation. 

Passarino-Veltman integrals can be computed numerically, but approximate analytical expressions can be used, as well.
The analytic solution for the scalar integral can be presented as 
\begin{equation}
\label{eqn:scalar-integral-analytical}
    \begin{aligned}
        &C_0(\{k_i,m_i\},\{k_j,m_j\})
         \approx\frac{1}{4(k_i\cdot k_j)}[\ln^2\frac{2(k_i\cdot k_j)}{m_j^2}\\&+(2\ln\frac{2(k_i\cdot k_j)}{m_j^2}+\ln\frac{m_j^2}{m_i^2})\ln\frac{m_j^2}{\lambda^2}-\frac{1}{2}\ln^2\frac{m_j^2}{m_i^2}\\&-2Li_2(-\frac{m_j^2-2(k_i\cdot k_j)}{2(k_i\cdot k_j)})]
\end{aligned}
\end{equation}
under the approximation $m_i\ll m_j$ \cite{kuraev&bytev&bystritskiy&gustafsson}, which significantly simplifies the evaluation for electron-hadron scattering. Here, the dilogarithm function is defined as $Li_2(z)=\int_z^0 \frac{\ln(1-t) dt}{t}$.

LETs \cite{low1954scattering, gell1954scattering,low1958bremsstrahlung, saito1969low} are crucial for understanding how scattering amplitudes or cross-sections behave under conditions of low energy or momentum during particle interactions. These theorems leverage fundamental symmetries, such as gauge symmetries, to predict outcomes reliably when interactions occur at low energy scales. Concurrently, in the context of TPE, the concept of soft photons becomes particularly relevant. Soft photons are characterized by their relatively low energy or momentum compared to the overall energy scale of the scattering process. They may emerge due to the emission or absorption of real or virtual photons during the interaction.

Soft photon emission is notably linked to infrared divergences, a common feature in quantum field theories dealing with electromagnetic interactions. In scattering processes, the emission of soft photons is essential for compensating the infrared divergences that arise from the exchange of virtual photons. This compensation ensures the infrared finiteness of physical observables, maintaining the consistency and predictability of the physical model.

Another important property of soft-photon exchange is that it involves low-momentum (or long-distance) scales that would not be able to resolve short-range quark structure of a nucleon. For this reason, separation of soft and hard momentum scales is necessary for TPE calculations. Proceeding further, we will use the approach implemented in the partonic calculation of TPE \cite{afanasev2005two}, with hard TPE included within a parton model, while including soft photons in TPE at a hadronic level. In application to SIDIS, it can be done by replacing charges and momenta of the final quark and the diquark by the charge and four-momentum of the detected hadron and the undetected spectator system, respectively.
%In this framework, soft photons, facilitated by the principles articulated in the LET, play a pivotal role in mitigating infrared divergence, thus securing the accuracy and stability of predictions at low energy scales.

\subsection{Bremsstrahlung process}
\begin{figure}[hbt!]
	\centering
	\includegraphics[scale=0.15]{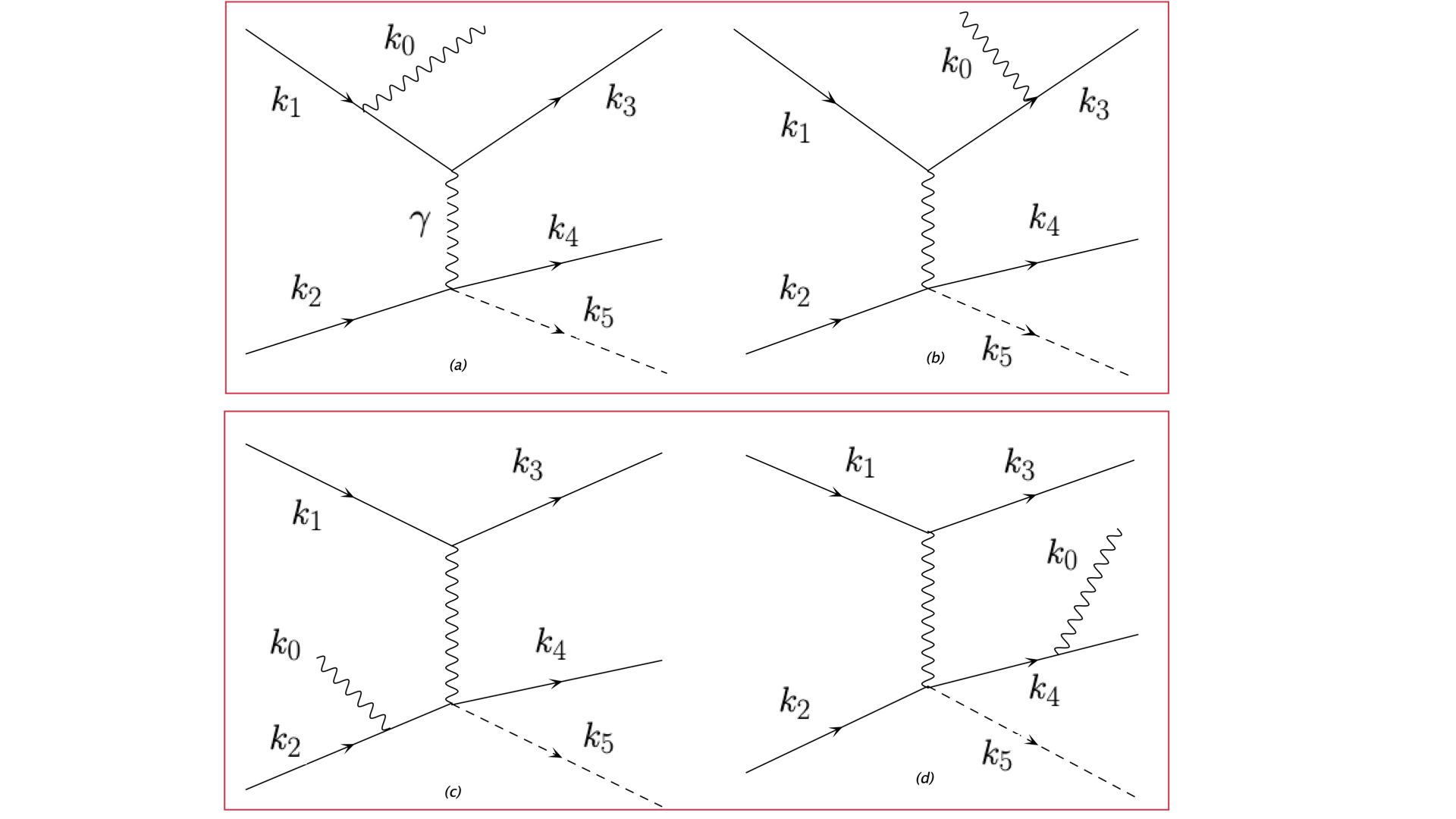}
	\vspace{-1em}
	\caption[Bremsstrahlung process]{The Bremsstrahlung process, where the upper two plots (figures (a) and (b)) correspond to Bethe-Heitler process, and the bottom two plots (figures (c) and (d)) correspond to Born-level virtual Compton scattering (VCS).}
	\label{fig:bremsstrahlung}
\end{figure}

In the second part of the calculation, the correction for inelastic scattering is considered, as illustrated in Fig.\;\ref{fig:bremsstrahlung}. When calculating the matrix element, the polarization vector of the emitted photon, $\epsilon^{*}$, must be included. The vertex function, which connects the real photon and the proton current \cite{afanasev&aleksejevs&barkanova}%\cite{gangadharan2023bremsstrahlungcrosssectionpolarized}
, allows the amplitude to be expressed as:
\begin{equation}
    M_{\gamma}^{(\pi)}=M^{1\gamma}\cdot e_i\frac{k_4\cdot \epsilon^{*}(k_0)}{k_4\cdot k_0}
\end{equation}
with $e_i$ is the charge for different particles and $\sum_\epsilon[k_i\epsilon(k_0)][k_j\epsilon^*(k_0)]=-(k_i\cdot k_j)$. Summing over all relevant cases, the correction due to the Bremsstrahlung process is given by:
\begin{equation}\label{eqn:brem}
        \begin{split}
            \delta_\gamma\propto[-(k_1\cdot k_2)I(k_1,k_2)+(k_3\cdot k_2)I(k_3,k_2)\\-(k_3\cdot k_4)I(k_3,k_4)+(k_1\cdot k_4)I(k_1,k_4)]
        \end{split},
\end{equation}
where the soft-photon integrals
\begin{equation}
    \begin{aligned}
        I(k_i,k_j)&=\int \frac{d^3\mathbf{k_0}}{\sqrt{\mathbf{k_0}^2+\lambda^2}}\frac{1}{(k_i\cdot k_0)(k_j\cdot k_0)}
    \end{aligned}
\end{equation}
 with indexes $i,j=1,2,3,4$ as the momenta from Fig$.$ \ref{fig:bremsstrahlung}, $k_0$ denoting the momentum of the virtual photon, and $\lambda$ being a fictitious photon mass introduced to regularize the IR divergence. 
 
 Upon evaluating the integral, $I(k_i, k_j)$ includes a logarithmic term of the form $\ln\left( \frac{4\Delta\epsilon^2}{\lambda^2} \right)$, where the parameter $\Delta\epsilon = \frac{m_4(m_5 + \Lambda)}{2\Lambda}$ defines the upper limit of the soft-photon energy. Here, $\boldsymbol{\Lambda} = \mathbf{k}_5$ is the momentum of the recoiling final-state particle.

 The dependence on the fictitious photon mass $\lambda$ cancels the IR divergence arising from the corresponding three-point scalar loop integral. For maximum soft-photon energy, we choose the frame where $\boldsymbol{\Lambda} = \mathbf{0}$, commonly referred to as the $\mathbf{R}$ frame. In this frame, the maximum value of $\Lambda \equiv \Lambda_{\text{max}} = m_4 + m_5$, and the product $m_4(m_5 + \Lambda)$ simplifies to $\Lambda_{\text{max}}^2 - m_5^2$.

 Moreover, $\Lambda_{\text{max}}$ can also be expressed as $W - m_4$, where W is the invariant mass of the system composed of the virtual photon and the incoming particle. As a result, the maximum value of $\Delta\epsilon$ can be written as
\begin{equation}
    \Delta\epsilon_{max}=\frac{(W-m_4)^2-m_5^2}{2(W-m_4)}.
\end{equation}
Additionally, the missing mass during the bremsstrahlung process, $\nu=W^2+m_4^2-m_5^2-2WE_{quark}$, is needed to be considered due to the emission of a bremsstrahlung photon. Because of the threshold of electroproduction, the largest possible value for the mass, $\nu_{max}$, such that $m_5^2=(W-m_4)^2-\nu_{max}$ when $E_{quark}=m_4$ and $\nu_{max}$ is always smaller than the heavier hadron. 

Overall, the final result of the total correction for quark-diquark model will be $\delta^{TPE}=\delta_{box}^{SPT}+\delta_\gamma$.

\subsection{Azimuthal Moments $\braket{\cos{(n\phi)}}$}	

Azimuthal asymmetries emerge when the hadron emitted in SIDIS is not collinear with the virtual photon, for instance, when the transverse momentum of the hadron is nonzero. These asymmetries can be parameterized using the cross-section decomposition in terms of structure functions using a one-photon exchange approximation \cite{gourdin1972semi,bacchetta&boer&diehl&mulders}:
\begin{equation}
    d\sigma_0=\epsilon\cdot\sigma_L+\sigma_T+\sqrt{2\epsilon(\epsilon+1)}\sigma_{LT}\cos\phi+\epsilon\cdot\sigma_{TT}\cos2\phi 
\end{equation}
Here, $\sigma_0$ corresponds to the cross-section in one-photon exchange approximation that allows model-independent separation of electron's kinematic parameters $\phi$ and $\epsilon$. The term $\cos{\phi}$ arises from interference between the transverse and longitudinal components of the electromagnetic current (with respect to the virtual photon's direction), while the term $\cos{2\phi}$ originates from interference between transverse current components. Understanding this azimuthal dependence is crucial for probing the underlying reaction mechanisms and extracting relevant physical observables.
 In particular, the $\braket{\cos{\phi}}$ and $\braket{\cos{2\phi}}$ moments are linked to the Cahn effect \cite{Cahn1978} and Boer-Mulders TMD \cite{BOER2003201}, respectively, both of which reflect correlations involving the intrinsic transverse momentum of quarks and their spin. Therefore, precise measurements of azimuthal asymmetries offer a valuable means to extract and constrain TMDs from experimental data, see Ref.\cite{boussarie2023tmd} for further discussion of TMDs in the context of SIDIS observables.

This study investigates the effect of TPE corrections on the dependence of SIDIS cross sections on the azimuthal angle ($\phi$) of the detected meson that is expressed through \(\braket{\cos(n\phi)}\) moments, as well as TPE contribution to extraction of $R=\sigma_L/\sigma_T$ ratio. Using kinematics for upcoming Jefferson Lab experiments \cite{E12-06-104,capobianco2023measurements,JLAB-E12-06-112}, we assess the importance of these corrections in determining azimuthal asymmetries, specifically \(\braket{\cos{\phi}}\) and \(\braket{\cos{2\phi}}\), for future experimental measurements.

The relation between the cross-sections and the asymmetric factors can be written as 
\begin{equation}
    \begin{split}
       &\frac{d\sigma_{tot}}{dxdzdQ^2d^2P_T}\equiv d\sigma_{tot}\\&=d\sigma_{exp}/(1+\delta^{TPE})
       =(1-\delta^{TPE})d\sigma_{exp}\\
       &\sim(1-\delta^{TPE}) \{K(y)[(1+\epsilon \frac{F_{UU,L}}{F_{UU,T}})\\&+\sqrt{2\epsilon(1+\epsilon)}\cos{2\phi}\frac{F_{UU}^{\cos{(2\phi)}}}{F_{UU,T}}
	    +\epsilon\cos{\phi}\frac{F_{UU}^{\cos{\phi}}}{F_{UU,T}}]\}
    \end{split}
    \label{eqn:cross-section}
\end{equation}
$F_{UU,X}$ are structure functions, and the kinematic variables are defined as 
\begin{align}
    K(y)&=1-y+y^2/2+\gamma^2 y^2/4\\
    \epsilon&=\frac{1-y-\gamma^2 y^2}{K(y)}\\
    \gamma&=2m_2x_{BJ}/Q\\
	P_T&=p_{4}\sin(\theta_{4})\\
	x_{BJ}&=\frac{Q^2}{2k_2\cdot q}\\
    y&=\nu_E/E=\frac{k_2\cdot q}{k_2\cdot k_1}\\
	z&=\frac{k_2\cdot k_4}{k_2\cdot q}\\
    \nu_E&=E_{lab}-E'\\ \nonumber
\end{align} 
with the invariant mass W, transverse momentum $P_T$, $E_{lab}$ is the incoming beam energy, $E'$ is the energy of scattered electron, $p_4$ is the momentum of quark, $\theta_4$ is the angle between virtual photon and $p_4$. 
After simplification, the $\braket{\cos{(n\phi)}}$ moments with correction can be interpreted as 

\begin{equation}
        \braket{\cos{(n\phi)}}\sim\int d\sigma_{exp}d\phi (1-\delta^{TPE})\cos{(n\phi)}
\end{equation}
with $\phi$ dependent correction $\delta^{TPE}$. $\theta_4$ and $\phi$ are the polar and azimuthal angles of the detected quark (diquark). In order to see how significant the corrections can affect the cosine moments, the results of $\braket{\delta\cos{(n\phi)}}\sim\int d\sigma_{exp}d\phi \delta^{TPE}\cos{(n\phi)}$ will also be presented. 

Because TMDs are extracted through precise comparisons between measured azimuthal modulations of the cross sections and theoretical predictions ($e.g.$, \cite{bacchetta2022unpolarized}), after applying ``standard" radiative corrections \cite{akushevich1999radiative, akushevich2009lowest, afanasev2024radiative}, quantifying the impact of TPE corrections is critical for reducing potential systematic uncertainties. Even small distortions at $\braket{\cos{(n\phi)}}$ moments can bias the interpretation of TMDs. One may recognize that if the ratio $R=\sigma_L/\sigma_T=\frac{F_{UU,L}}{F_{UU,T}}$ is nonzero, then within one-photon-exchange approximation it can be extracted using Rosenbluth separation as a coefficient in front of kinematic variable $\epsilon$. However, if the TPE correction is also dependent on $\epsilon$, the experimentally measured cross section would show a nonzero $\epsilon$ slope even if $R=0$.  Physics implications of observing $R$ are discussed in Ref.\cite{}The numerical example below will show that TPE corrections to the measurements of $R$ are important.

Therefore, understanding how TPE effects alter SIDIS observables is essential for TMD phenomenology.

\section{Results and Discussion}
This section presents the results for TPE calculations if the detected mesons are $\pi^+$ ($m_\pi=0.14$ GeV), %represented by blue solid lines in the figures) 
or $\rho^+$ ($m_\rho=0.77$ GeV).
%represented by orange dashed lines),
As was elaborated above, for the calculation of soft-photon contribution into TPE, the charge and mass of the quark is replaced by the charge and mass of corresponding hadrons ($e_q=1$), while the diquark mass and charge are chosen to match the properties of the spectator system of hadrons; in the plots, we choose
the mass of the spectator $m_x=m_5=2$ GeV and $e_S=0$. For this example, only positively charged mesons are considered in this analysis, while extension to other charges is straightforward. As such, the calculations involving the ``diquark graph" discussed earlier are excluded. Kinematic variables  were set as follows: incoming beam energy, $E_{lab}=10.6$ GeV; $Q^2=2.5$~GeV$^2$; the mean value of the Bjorken-x, $x_{BJ}$, is 0.31 (corresponding to the invariant mass $W\approx 2.7$ GeV); the kinematic range of {\it y} was maintained below 0.75 to minimize susceptibility to radiative effects and lepton-pair symmetric backgrounds \cite{E12-06-104,capobianco2023measurements}; $z=0.7$. The polar angle of the detected mesons is set such that $\cos(\theta_4)=0.8$ (corresponding to the transverse momentum $P_T\approx 0.4$ GeV$/c$), suitable for $P_T$-independent figures, and the azimuthal angle is fixed at $\phi=\pi/6$ for $\phi$-independent plots.

The ratio of the structure functions anticipated for Jefferson Lab kinematics \cite{E12-06-104,capobianco2023measurements,JLAB-E12-06-112} in Eqn.\;\ref{eqn:cross-section} are as follows: 
\begin{align}
	&F_{UU,L}/F_{UU,T}\approx 0.2 \nonumber \\ 
    &F_{UU}^{\cos{\phi}}/F_{UU,T}\approx -0.05\\
    &F_{UU}^{\cos{(2\phi)}}/F_{UU,T}\approx 0.1, \nonumber
    \label{eq:SFratios}
\end{align} 
that in one-photon exchange approximation would describe dependence of the SIDIS cross section on the final meson's azimuthal angle $\phi$ and electron scattering angle (via the variable $\epsilon$). The measurements from HERMES \cite{airapetian2013azimuthal} and COMPASS \cite{adolph2014measurement, compassnpb2020} indicate that the cosine moments for \(\braket{\cos{\phi}}\) are typically in the range of approximately $-0.05$ to $0$, and from $0$ to $0.05$ for \(\braket{\cos{2\phi}}\). Note that none of the above-mentioned experiments considered TPE in the data analysis. Consequently, the interpretation of the extracted TMDs may be subject to systematic biases if these corrections are neglected, particularly in regions of moderate $Q^2$ and sizable $P_T$ where TPE effects are not negligible.

\begin{figure}[H]
\begin{minipage}[]{1\linewidth}
    \centering
    \includegraphics[scale=0.4]{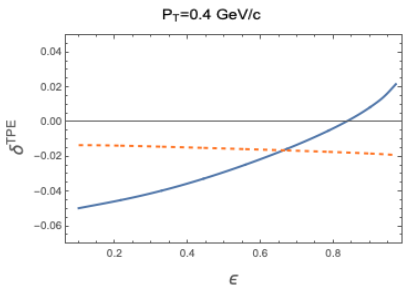}\\
    \label{}
\end{minipage}
\\
\vspace{0.2 cm}
\begin{minipage}[]{1\linewidth}
    \centering
    \includegraphics[scale=0.4]{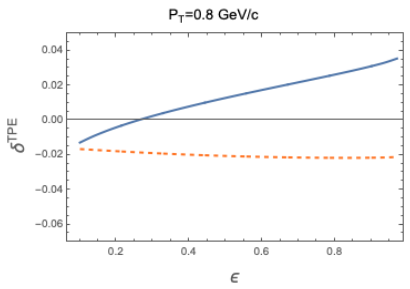}\\
    \label{}
\end{minipage}
\\
\vspace{0.2 cm}
\begin{minipage}[]{1\linewidth}
    \centering
    \includegraphics[scale=0.4]{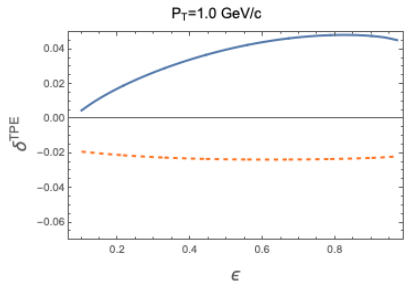}\\
    \label{}
\end{minipage}
\caption{\small{TPE correction $\delta^{TPE}$ as a function of $\epsilon$ for fixed values of transverse momentum $P_T=0.4,0.8, 1.0$ (GeV$/c$) at $z=0.7$, $Q^2=2.5$~GeV$^2$, $x=0.31$, and $\phi=\pi/6$. The blue solid line and the orange dashed line indicate the detected meson is $\pi^+$ and $\rho^+$, respectively. }}
\label{fig:delta-epsilon-pt} 
\end{figure}

\begin{figure}[H]
    \centering
	\includegraphics[scale=0.29]{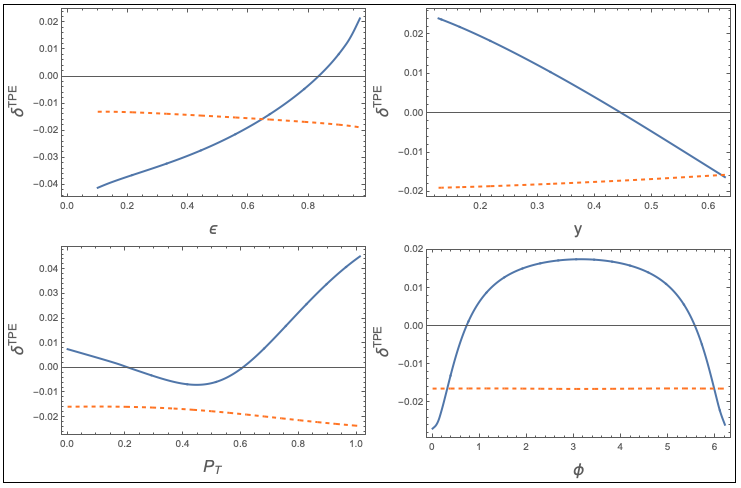}
	\caption{\small{Dependence of TPE correction $\delta^{TPE}$ on the virtual photon $\epsilon$, electron's relative energy loss $y$, transverse momentum $P_T$, and azimuthal angle $\phi$ with $E_{lab}=10.6$ GeV, $Q^2=2.5$ GeV$^2$, the mean value $\braket{x_{BJ}}=0.31$,  $z=0.7$, using kinematics for projected experiments \cite{E12-06-104,capobianco2023measurements}. The blue solid line and the orange dashed line represent the detected mesons are $\pi^+$ and $\rho^+$ meson, respectively. }}
	%\small{\hspace{1.5cm}(a)}
    \label{fig:tot-delta}
\end{figure}

Next, we evaluate effects from TPE corrections. Fig.\;\ref{fig:delta-epsilon-pt}-\ref{fig:tot-delta}  illustrate the role of the TPE corrections ($\delta^{TPE}$) for ranges of  kinematic variables $\epsilon$, $y$, transverse momentum $P_T$, and the azimuthal angle $\phi$. The data indicate that the intervals of corrections for the $\pi^+$ vary approximately between $-5\%\sim 5\%$, while those for the $\rho^+$ meson falls within about $1\%$. 
%As shown, the value of transverse momentum increases as the lower end of the corrections rises, with the maximum values of the corrections remaining relatively stable. 
Comparing TPE corrections for detected $\pi^+$ mesons vs $\rho^+$ mesons,  we note smaller TPE corrections if the detected meson's mass is larger. We emphasize that the leading TPE effect arises from the correction to the largest $\sigma_T$ term ($i.e.$ to the structure function $F_{UU,T})$. This means that numerical results for TPE presented here have negligible dependence on the values of other structure functions (as long as they are much smaller than $F_{UU,T} $), and the TPE correction is {\it additive} - and not {\it multiplicative} - when evaluating $\cos(n\phi)$ moments or $\sigma_L/\sigma_T$ ratios. 

\begin{figure}[H]
    \centering
	\includegraphics[scale=0.29]{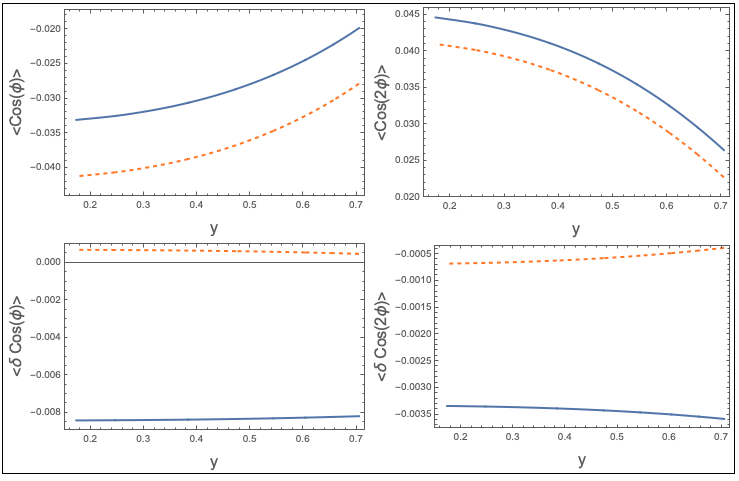}
	\caption{\small{The cosine moments as a function of variable $y$ when $P_T= 0.4$ GeV$/c$ and other kinematic values remain the same. Kinematics is as in Jefferson Lab experiments Refs.\cite{E12-06-104,capobianco2023measurements}. The blue solid line and the orange dashed line represent the detected mesons are $\pi^+$ and $\rho^+$ meson, respectively. }}
	\label{fig:cos-y}
\end{figure}

\begin{figure}[H]
%\begin{minipage}[]{1\linewidth}
    \centering
    \includegraphics[scale=0.29]{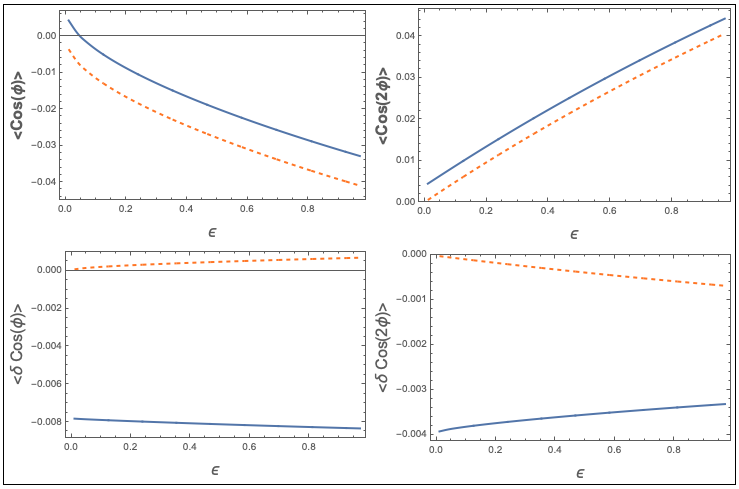}
	\caption{\small{The cosine moments as a function of variable  $\epsilon$ when $P_T\approx 0.4$ GeV$/c$ and other kinematic values remain the same. Kinematics is from  Refs.\cite{E12-06-104,capobianco2023measurements}. The blue solid line and the orange dashed line represent the detected mesons are $\pi^+$ and $\rho^+$ meson, respectively. }}
	\label{fig:cos-epsilon}
%\end{minipage}
\end{figure}

Fig.\;\ref{fig:cos-y}-\ref{fig:cos-epsilon} depict the relationships between the cosine moments and different kinematic variables such as $y,\epsilon, P_T$. The trends for the correction-only terms, $\braket{\delta\cos{(n\phi)}}$, are consistent with those illustrated in Fig.\;\ref{fig:delta-epsilon-pt}-\ref{fig:tot-delta}, depending on the masses of the detected mesons. However, the trends for the overall corrected moments for $\pi^+$ and $\rho^+$ meson, $\braket{\cos{(n\phi)}}$, exhibit less than $0.5\%$ difference across the respective kinematic ranges.

We have not presented results for $n=3$ in $\cos{n\phi}$ even though they are nonzero if TPE is included. However, they are significantly smaller in magnitude compared to those for $n=1\;\&\;2$; results for $\braket{\cos{3\phi}}$ are around $0.2\%$ of $\braket{\cos{2\phi}}$ regardless of dependences on either $y$, $\epsilon$, or $P_T$, and those for $\braket{\delta\cos{3\phi}}$ are even smaller, of the order of $10^{-7}$. 

Jefferson Lab has proposed experiments \cite{JLAB-E12-06-112} that focus on extracting the longitudinal virtual photon contributions to SIDIS, and there are yet to be any measurements available to evaluate systematics from the TPE. The findings presented above underscore the relevance of considering TPE corrections in experimental setups to enhance our understanding of the underlying structure of hadrons. This may include accessing transverse momentum distributions, constraining quark and gluon polarizations, and testing predictions from Quantum Chromodynamics (QCD). The overall results demonstrate that these corrections are a few percent for different kinematic dependencies, a factor worth considering in experimental analyses.

\begin{figure}[H]
%\begin{minipage}[]{1\linewidth}
    \centering
	\includegraphics[scale=0.29]{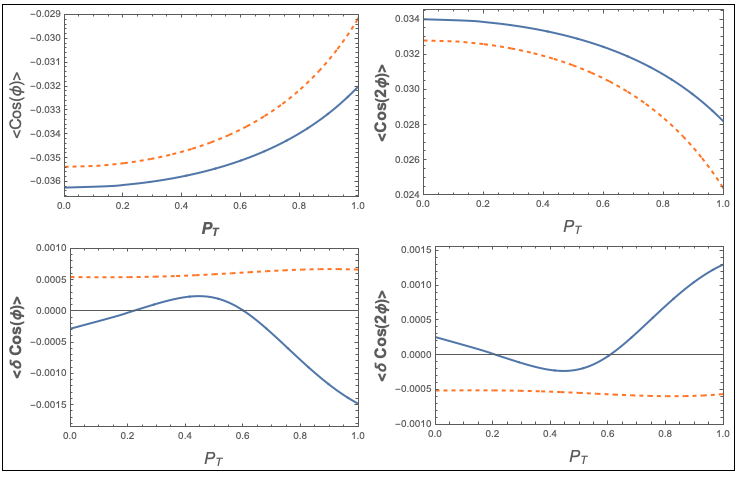}
	\caption{\small{The cosine moments  as a function of transverse momentum of the detected meson. Kinematics is from Refs. \cite{E12-06-104,capobianco2023measurements}. The blue solid line and the orange dashed line represent the detected mesons are $\pi^+$ and $\rho^+$ meson, respectively. }}
	\label{fig:cos-pt}
%\end{minipage}
\end{figure}

\section{Summary}
This study investigated the role of TPE corrections in SIDIS, focusing on their effects on structure function ratios, azimuthal asymmetries, and overall cross-section modifications. Using the soft-photon approximation, we calculated the TPE corrections for positively charged pions ($\pi^+$) and rho mesons ($\rho^+$), highlighting the dependence of these effects on meson mass and various kinematic conditions. Our results indicate that TPE corrections typically fall within a few percent, with cross sections for heavier mesons experiencing smaller corrections. Moreover, we show that TPE corrections depend on the hadronic transverse momentum, thereby influencing extraction of TMDs from observables.

The $\phi$-modulated azimuthal asymmetries, used to extract TMDs such as the Boer–Mulders function ($h_1^\perp$) 
%Sivers function ($f_{1T}^\perp$), 
and unpolarized TMD ($f_1(x, k_T)$), can be distorted by two-photon exchange effects. Our calculations show that higher-order cosine moments such as \(\braket{\cos{3\phi}}\) are non-zero but negligible compared to \(\braket{\cos{\phi}}\) and \(\braket{\cos{2\phi}}\), reinforcing the dominance of lower-order angular modulations in SIDIS processes. Since the extraction of TMDs heavily relies on fits to these angular moments, any unaccounted TPE corrections may introduce systematic errors.

TPE correction is thus an additional systematic effect that must be included in the data analysis of future SIDIS measurements, as well as in the interpretation of previous experiments in terms of TMDs. Given our estimates for TPE found in the 5\% range, the experiments at HERMES \cite{airapetian2013azimuthal} and COMPASS \cite{adolph2014measurement, compassnpb2020}, which measured the moments for \(\braket{\cos{\phi}}\) in the range of approximately $-0.05$ to $0$, and from $0$ to $0.05$ for \(\braket{\cos{2\phi}}\), could be significantly affected by TPE systematics if not properly corrected.

By employing the soft-photon approach, we provided a model-independent estimate of the TPE corrections, ensuring that our calculations adhered to LETs. This methodology allowed us to isolate the leading TPE effects without requiring detailed assumptions about the internal structure of the nucleon, making the results broadly applicable to SIDIS experiments. However, a complete description of TPE effects would require incorporating hard-photon exchange contributions, which involve explicit quark-level calculations.

Our findings underscore the importance of including TPE corrections in precision SIDIS analyses, particularly for future precision experiments at facilities such as Jefferson Lab and EIC. As ongoing and future experiments aim to extract longitudinal photon contributions and refine our understanding of TMDs, properly accounting for TPE effects will be essential for reducing systematic uncertainties and improving theoretical interpretations.

Future studies could extend this work by incorporating higher-order TPE contributions and examining their impact within different factorization frameworks. Furthermore, direct experimental validation of the TPE effects in SIDIS remains an open challenge. 
%Such measurements would provide valuable insights into the electromagnetic structure of nucleons and help constrain QCD-based models of hadronic interactions. Ultimately, a deeper understanding of TPE in SIDIS will enhance our ability to probe nucleon structure, refine parton distribution functions, and test fundamental predictions of QCD in the non-perturbative regime. 
It can be addressed by using positron scattering in comparison to electron scattering. This possibility is being considered at Jefferson Lab \cite{accardi2021experimental}.

\section*{Acknowledgments}
We gratefully acknowledge useful discussions with Harut~Avakian. This work was supported by National Science Foundation award PHY-2111063.
%\clearpage
%\nocite{*}
\bibliography{citation}

\end{document}